\documentclass[aps,prl,reprint,groupedaddress]{revtex4-1}

\usepackage{graphicx}
\usepackage{bm}

\newcommand{\sfbold}{\bf \sffamily}

\begin{document}

\title{Detection of the phase shift from a single quantized superconducting vortex}

\author{Taras Golod, Andreas Rydh \& Vladimir M. Krasnov}
\thanks{Electronic address: vladimir.krasnov@fysik.su.se}

\affiliation{
Department of Physics, Stockholm University, AlbaNova University Center, SE~--~106 91 Stockholm, Sweden}

\begin{abstract}
An Abrikosov vortex in a superconductor carries a flux quantum, $\bm{\Phi_0 = h c/2e}$, localized at its center, but induces a global $\bm{2 \pi}$ phase rotation in the superconducting condensate. This long-range gauge field\cite{Wu} outside the area pierced by a magnetic field is due to the Aharonov-Bohm effect\cite{AB}, which is a non-classical phenomenon that illustrates the significance of potentials rather than forces in quantum mechanics. In the London gauge, the phase of the condensate is given by the polar angle around the vortex. Here we raise the question whether this phase shift could be detected by means of Cooper pair interferometry using Josephson junctions as phase-sensitive detectors. We introduce a single Abrikosov vortex into a superconducting lead with a detector junction made at the edge of the lead. We observe that the vortex induces a Josephson phase shift equal to the polar angle of the vortex within the junction length. When the vortex is close to the junction it induces a $\bm{\pi}$-step in the Josephson phase difference, leading to a controllable and reversible switching of the junction into the $\bm{0}$~--~$\bm{\pi}$ state\cite{Bul2,VHarlingen,Tsuei,Hilgenkamp,Ryazanov,Aprili, Baselmans, Weides, Gaber, Frolov}. This in turn results in an unusual $\bm{\Phi_0/2}$ quantization of the flux in the junction. The vortex may hence act as a tunable ``phase battery" for quantum electronics.
\end{abstract}

\maketitle

\begin{figure}
\begin{center}
\includegraphics[width=0.95\linewidth]{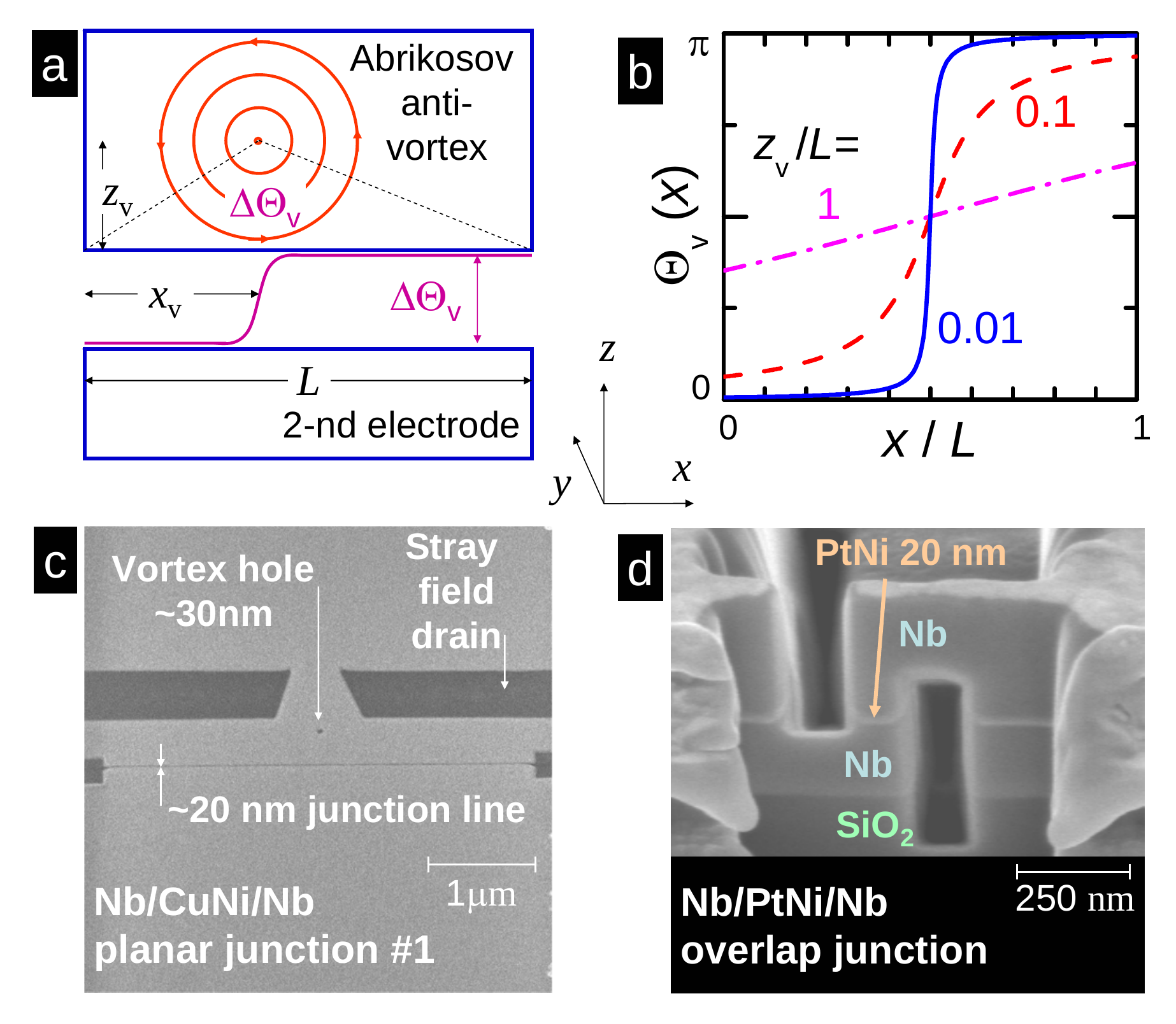}
\end{center}
\caption{{\sfbold Experimental configuration and sample geometry.} {\sfbold a}, Geometry of the experiment: the phase shift from the single Abrikosov vortex is detected by the junction at the edge of the lead. The polar angle of the vortex within the junction length $\Delta \Theta_v$ is marked by the dashed lines. {\sfbold b}, The polar angle of the vortex $\Theta_v(x)$ along the junction length for different distances $z_v$ from the vortex to the junction and $x_v=L/2$. {\sfbold c}, Top view of a planar Nb/CuNi/Nb junction $\# 1$, with a vortex hole and a stray field drain. {\sfbold d}, SEM image of a nano-sculptured Nb/PtNi/Nb junction. SEM images in c and d are shown in the same perspective as the sketch in a. The magnetic field in our experiment is applied along the $y$-axis (into the paper).
\label{Fig1}}
\end{figure}

A sketch of our experiment is shown in Fig.~\ref{Fig1}a. A single Abrikosov vortex is placed in a superconducting lead. A detector Josephson junction is made at the edge of the lead. The supercurrent $I_c$ through the junction is a result of interference of Cooper pair wave functions, which leads to a Fraunhofer modulation of $I_c$  as a function of magnetic field. The Josephson phase shift, induced by the vortex, is detected from a comparison of $I_c(H)$ patterns with and without the vortex.

In the London gauge, variation of the phase of the superconducting condensate around the vortex is given by the polar angle $\Theta_v$, which, at the junction interface is equal to:
\begin{equation}
\Theta_{v}(x) = \arctan \left(\frac{x-x_v}{z_v}\right) + \mathrm{const.},
\label{Q1}
\end{equation}
where $x_v$ and $z_v$ are the vortex coordinates and $x$ is the position along the junction length. Profiles of $\Theta_v(x)$ for different distances from the vortex to the junction are shown in Fig.~\ref{Fig1}b. Even in quantum mechanics gauge fields have limited physical significance. Only closed path integrals of gauge fields are measurable\cite{Wu}. For Cooper pairs such integrals around the vortex are equal to $2\pi$, which is indistinguishable from $0$ in the absence of the vortex. Open path integrals are not gauge-invariant and should not be measurable. Therefore, the question whether a distant Abrikosov vortex gives rise to a Josephson junction phase shift is non-trivial.

The main challenge for the present experiment is to avoid vortex intrusion into the junction area, which might induce a parasitic phase shift in the junction. Conventional Josephson junctions are formed by a barrier sandwiched between thin superconducting films. Abrikosov vortices in such ``overlap" junctions tend to minimize their energy by orienting themselves perpendicular to the electrodes, thus introducing a segment of a Josephson vortex (fluxon) in the junction. This is a well-known reason for distortion of $I_c(H)$ patterns in conventional overlap junctions subjected to out-of-plane fields\cite{Golubov,Lisitskii}. To avoid fluxon formation we employ two types of specially designed detector junctions:

\begin{enumerate}
\item \emph{Planar Nb/CuNi/Nb junctions}, see Fig.~\ref{Fig1}c. Such junctions are ideal for the planned experiment: due to their two-dimensional geometry, the Abrikosov vortex, which is oriented perpendicular to the Nb film, cannot cross the junction \emph{line}.
\item \emph{Mesoscopic Nb/PtNi/Nb junctions}, see Fig.~\ref{Fig1}d. Mesoscopic sizes of the junctions confine the vortex in the middle of the electrode parallel to the junction plane\cite{MesoVort2} and allow the detection of $I_c$ in very strong magnetic fields (up to 20 kOe), which further helps to align the vortex.
\end{enumerate}

\begin{figure}
\begin{center}
\includegraphics[width=0.90\linewidth]{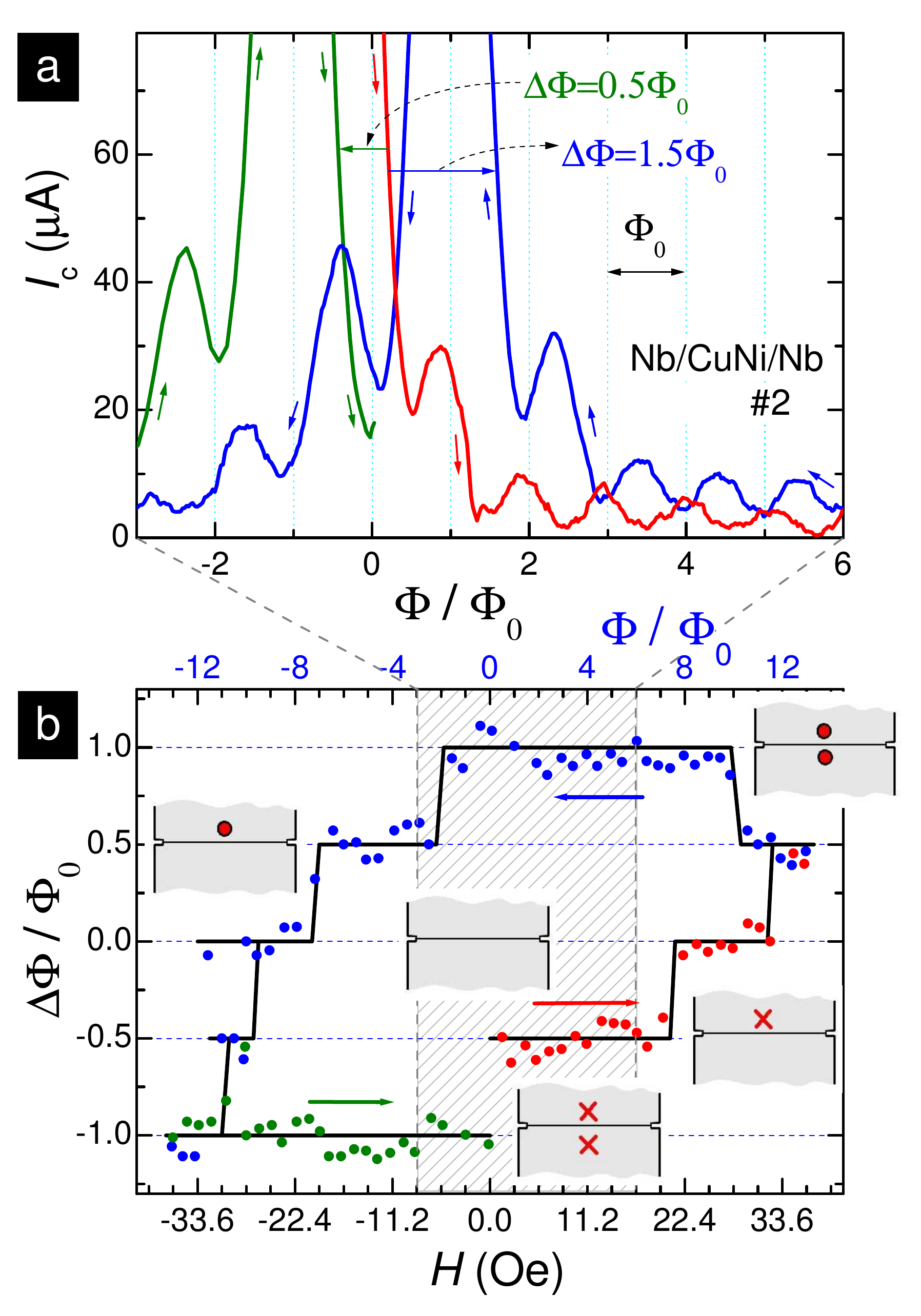}
\end{center}
\caption{{\sfbold Quantized flux offset in Fraunhofer patterns $\bm{I_c(\Phi)}$ of a planar junction without vortex hole or stray field drain.} {\sfbold a}, measured $I_c(\Phi)$ patterns for consecutive field sweeps from 0 to 40 Oe (red), from 40 to $-40$ Oe (blue) and from $-40$ to 0 Oe (green). The appearance of hysteresis (flux offset) upon sweeping the field is clearly seen. Note that the $I_c(\Phi)$ modulation for the first sweep (red curve) is out-of-phase with that on the way back (blue curve), which indicates that the offset is half-integer of $\Phi_0$. {\sfbold b}, Measured flux offset vs. $H (\Phi)$ for the same field loop (hatched area corresponds to the range shown in panel~a). Symbols represent minima or maxima in $I_c(\Phi)$. The unusual $\Phi_0/2$ quantization of $\Delta \Phi$ is clearly seen. Each step corresponds to a sequential entrance or exit of one Abrikosov vortex. The expected vortex configurations are indicated by adjacent sketches. Note that the offset of $I_c(\Phi)$ patterns occurs in the direction of the applied field, which implies that the trapped flux in the junction is opposite to the applied field.
\label{Fig2}}
\end{figure}

In Fig.~\ref{Fig2}a we show $I_c$ for a planar Nb/CuNi/Nb junction as a function of the magnetic flux $\Phi$ through the junction (field is oriented perpendicular to electrodes, along the $y$-axis in Fig.~\ref{Fig1}a). Measurements were done by first sweeping the field from $0$ to $40$~Oe, then to $-40$~Oe and finally back to $0$. The $I_c(\Phi)$ patterns are almost identical for all three sweeps, except for an offset $\Delta \Phi$, which changes stepwise with the field, as shown in Fig.~\ref{Fig2}b. The apparent quantization of $\Delta \Phi$ implies that each step is caused by entrance or removal of an Abrikosov vortex in the electrodes. Remarkably, the offset $\Delta \Phi$ is quantized in {\it half} flux quanta. As a result, the $I_c(\Phi)$ modulation gets out-of-phase, i.e., positions of minima and maxima are interchanged, with each step in $\Delta \Phi$. Furthermore, the offset occurs in the direction of applied field, which means that the trapped field in the junction is {\it opposite} to the direction of the field.

\begin{figure}
\begin{center}
\includegraphics[width=0.90\linewidth]{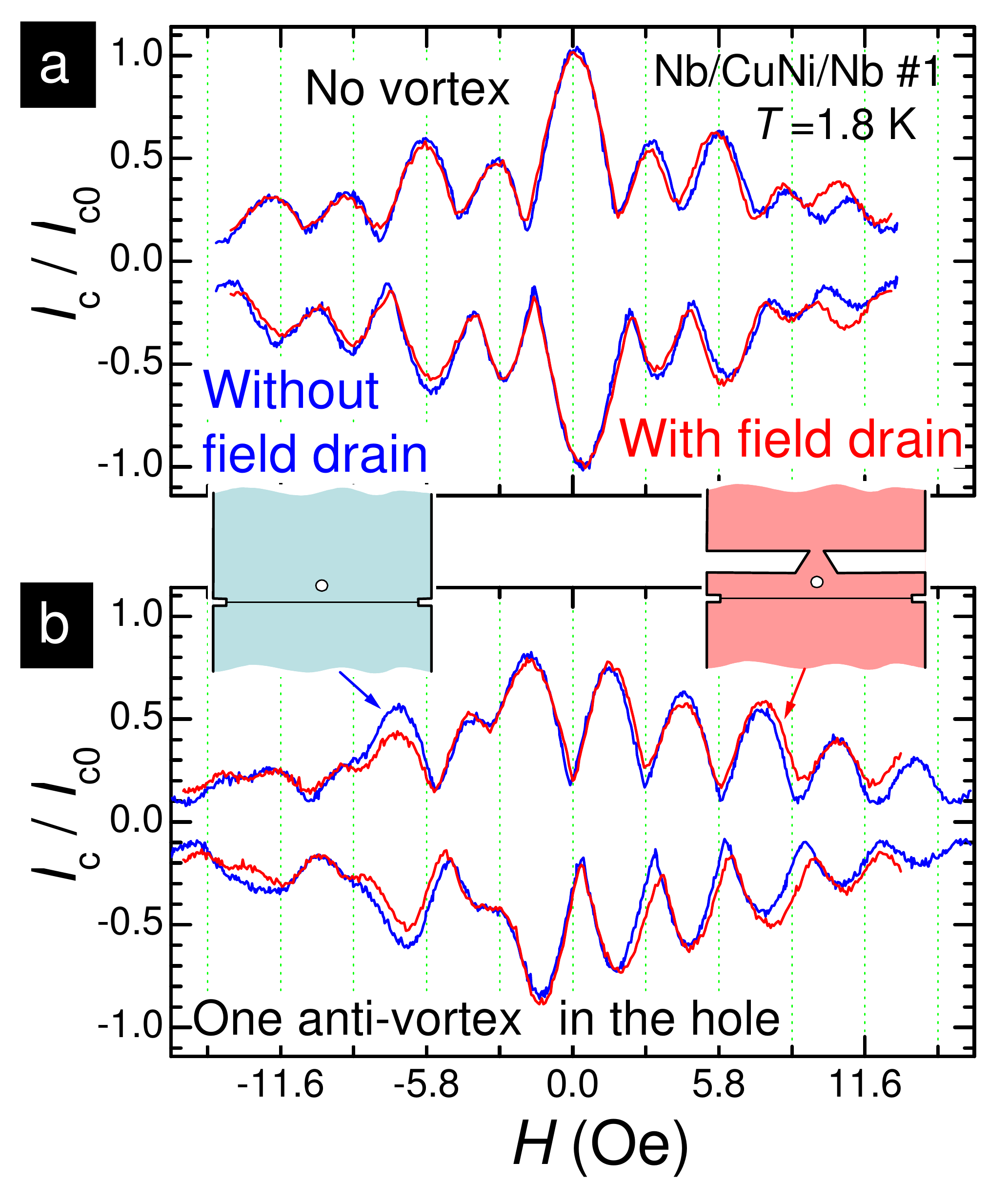}
\end{center}
\caption{{\sfbold $\bm{I_c(H)}$ modulation for a planar junction with the vortex hole and the stray field drain.} Sketches demonstrate junction sample geometries: blue curves - with the vortex hole only, red curves - with the hole and the stray-field drain. The current is normalized on the maximum critical current in the absence of vortices, $I_{c0}$. {\sfbold a}, Without vortices. {\sfbold b}, With an anti-vortex in the hole. To trap the anti-vortex, a sufficiently large negative field was applied before the measurements. Clear signatures of the $0$~--~$\pi$ state are seen: i) the central maximum is replaced by a minimum, ii) modulation of $I_c(H)$ gets out-of-phase with that without the vortex and iii) doubling of the periodicity occurs at one side of the pattern. Note that introduction of the stray field drain affects neither the $I_c(H)$ pattern, nor $\Delta \Phi$.
\label{Fig3}}
\end{figure}

To clarify the origin of the unusual $\Phi_0/2$ quantization, the junctions were modified by a Focused Ion Beam (FIB) in two steps, as shown in Fig.~\ref{Fig1}c. First, a vortex trap was made in order to control the position of the vortex. The trap is a small hole $\sim 30$~nm in the center of one of the electrodes near the junction. Second, a stray field drain was added by removing a substantial part of the electrode in the vicinity of the vortex hole. The drain should substantially decrease the magnetostatic stray fields from the vortex at the junction. Figure~\ref{Fig3}a shows the $I_c(H)$ patterns after these modifications in the absence of Abrikosov vortices. It is seen that the shape of the $I_c(H)$ pattern remained intact after introducing the field drain, implying that the junction uniformity was not affected by the drain.

A vortex in the hole can be controllably introduced or removed by applying an appropriate field and by using Lorentz force from the transport current (see supplementary information). Figure~\ref{Fig3}b shows a new type of $I_c(H)$ patterns, which appears after trapping the anti-vortex in the hole. It has three new characteristic features:
\begin{enumerate}
\item The central maximum at $H=0$ is replaced by a minimum.
\item The $I_c(H)$ modulation is out-of-phase with that for the vortex-free pattern, i.e., $\Delta \Phi \simeq \Phi_0/2$.
\item The periodicity of the $I_c(H)$ modulation doubles at the left side of the pattern, leading to a clear left-right asymmetry\cite{NonUn}. When a vortex is trapped in the hole (instead of an anti-vortex), the $I_c(H)$ pattern becomes mirror reflected with respect to the $H=0$ axis (not shown).
\end{enumerate}
These are the well known fingerprints of $0$~--~$\pi$ junctions, with a step-like $\pi$-shift in the Josephson phase difference within the junction\cite{Bul2,Weides,Gaber,Frolov}. Properties of $\pi$-junctions with negative Josephson coupling and $0$~--~$\pi$ junctions have attracted significant attention in recent years, both due to the interesting physics involved, and the potential for new applications. So far, three types of $0$~--~$\pi$ junctions were realized based on: i) the $d$-wave symmetry of the order parameter in high $T_c$ superconductors\cite{VHarlingen,Tsuei,Hilgenkamp}, ii) the oscillatory nature of the proximity induced order parameter in hybrid superconductor/ferromagnet junctions\cite{Ryazanov,Aprili,Weides,Frolov}, and iii) the phase shift by current injection into the junction\cite{Baselmans,Gaber}. Here we demonstrate that a conventional $0$-junction can be switched into the $0$~--~$\pi$ state by a single Abrikosov vortex, parallel to the junction plane.

\begin{figure}
\begin{center}
\includegraphics[width=0.90\linewidth]{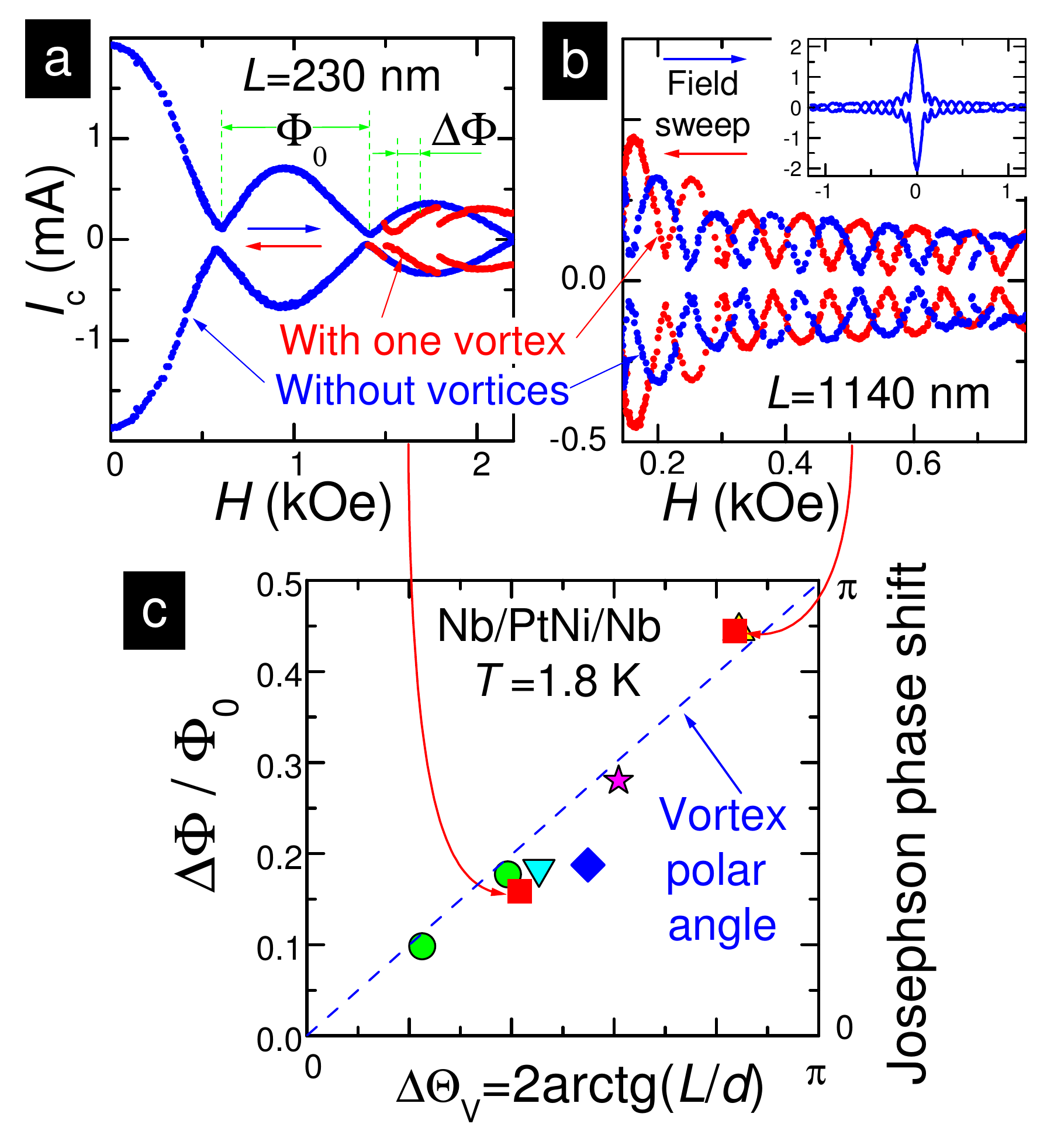}
\end{center}
\caption{{\sfbold Influence of the junction geometry on the flux offset in mesoscopic Nb/PtNi/Nb junctions.} {\sfbold  a}, {\sfbold b}, $I_c(H)$ patterns for the same Nb/PtNi/Nb junction $1140 \times 230$~nm$^2$ for two in-plane field orientations. First the field is swept up starting from the vortex free state until first Abrikosov vortices enter the electrodes (blue symbols) and then back to zero (red symbols). Entrance and exit of the Abrikosov vortex leads to a sudden appearance of the offset $\Delta \Phi$ in the Fraunhofer pattern. The inset in panel~b shows the $I_c(H)$ pattern in a wider field range. {\sfbold  c}, Flux offset $\Delta \Phi/\Phi_0$ induced by the vortex as a function of the polar angle $\Delta \Theta_v$ of the vortex within the junction. Similar symbols correspond to the same junction with different in-plane field orientations. Red squares correspond to the junction shown in panels~a and b. The dashed line indicates that the Josephson phase shift induced by the vortex (right axis) is simply equal to the polar angle of the vortex.
\label{Fig4}}
\end{figure}

Importantly, introduction of the stray field drain does not reduce the offset $\Delta \Phi$, see Fig.~\ref{Fig3}b. This clearly shows that the effective trapped flux in the junction is not driven by the simple magnetostatic spreading of the vortex stray field.
To get more insight into the influence of junction geometry on $\Delta \Phi$, mesoscopic Nb/PtNi/Nb junctions were used, where the vortex is geometrically confined in the middle of the electrodes\cite{MesoVort2}, $z_v \simeq d/2$, $x_v \simeq L/2$. All junctions had the same electrode thicknesses $d$, but different junction lengths $L$, thus allowing variation of the ratio $z_v/L \simeq d/2L$.

In Figs.~\ref{Fig4}a,b we show $I_c(H)$ for a Nb/PtNi/Nb junction $230 \times 1140$ nm$^2$ with an in-plane field parallel to different facets of the junction. The $\Delta \Phi$ induced by the vortex is clearly seen in both cases. For the long junction case, $L=1140$ nm, $\Delta \Phi \simeq \Phi_0/2$, which results in out-of-phase modulation of the patterns. However, for the short junction case, $L=230$ nm, $\Delta \Phi \simeq 0.16\Phi_0$ is considerably smaller.

Figure~\ref{Fig4}c summarizes the measured $\Delta \Phi $ for all studied Nb/PtNi/Nb junctions. Obviously, $\Delta \Phi$, introduced by the vortex increases considerably with the junction length $L$. The observed offset $\Delta \Phi$ in $I_c(H)$ corresponds to the net variation of the Josephson phase difference along the junction length
\begin{equation}
\Delta \varphi_v(L-0)= 2\pi \Delta \Phi/\Phi_0.
\label{dphi}
\end{equation}
Thus the Abrikosov vortex does induce a measurable Josephson phase difference in the junction. The dashed line in Fig.~\ref{Fig4}c indicates that the Josephson phase shift induced by the vortex is agreeing well with the polar angle of the vortex within the junction: $\Delta \varphi_v (L-0) \simeq \Delta \Theta_v$ (see Fig.~\ref{Fig1}a and equation~(\ref{Q1})). The $\Delta \Theta_v$ was calculated by assuming that the vortex is placed in the middle of an electrode $z_v\simeq d/2$, where $d\simeq 300$~nm is the average thickness of Nb electrodes.

To understand the origin of the observed phenomenon we first exclude several unsustainable scenarios:
\begin{enumerate}
\item
The vortex does not introduce a segment of a Josephson fluxon in the junction, which might distort $I_c(H)$\cite{Golubov,Lisitskii}. In our structures, the Abrikosov vortices were oriented strictly parallel to the junction interface and never cross the junction area. In particular, such crossing is impossible for the planar junctions, because of the two-dimensional junction geometry.
\item
A direct field of the vortex stretching into the junction or magnetism in the junction cannot explain the experiments, because the induced field in the junction is opposite to the applied field. Indeed, from Fig.~\ref{Fig2}a it is seen that the central maximum in $I_c(\Phi)$ is shifted to a positive field after applying $+40$~Oe and to a negative field after applying $-40$~Oe. Since the central maximum corresponds to zero total flux in the junction, the additional field within the junction is always {\it opposite} to the applied field.
\item
The magnitude of the signal can hardly be explained by a finite vortex current at the junction interface. Although it produces a phase shift of the proper sign, its magnitude should decay strongly with the distance from the vortex to the junction\cite{Aslamazov} and should for no reason produce a quantized $\Phi_0/2$ flux offset in the junction. Similarly, it is not possible to explain the characteristic equality between the Josephson phase shift and the polar angle of the vortex, shown in Fig.~\ref{Fig4}c.
\item
Magnetostatic stray field from the vortex would also give a phase shift with correct sign. The experiment with the stray field drain, however, demonstrates that the Josephson phase shift is unaffected by variation of magnetostatic conditions.
Furthermore, there is no reason for the stray field to be quantized as $\Phi_0/2$. When the vortex is placed very close to the junction, clear signatures of the $0$~--~$\pi$ junction are seen, see Fig.~\ref{Fig3}b. This implies that the induced Josephson phase shift has a form of a sharp $\pi$ step, which is again difficult to explain in terms of simple magnetostatics because it would require field focusing in one point. Additional discussion can be found in the supplementary.
\end{enumerate}

Our data show an unambiguous correspondence between the Josephson phase shift and the polar angle $\Theta_{v}$, which represents the variation of the phase of the superconducting condensate around the vortex within the London gauge, equation~(\ref{Q1}), as if the phase of the condensate is rigidly coupled to rotation of the current in the vortex. Numerical simulation presented in the supplementary demonstrate that equation~(\ref{Q1}) provides a good overall agreement with all our observations. It naturally explains the unusual half-integer flux quantization in the junction. The associated $\pi$-Josephson phase shift in this case is simply equal to the change in the polar angle upon going from the left to the right side of the Abrikosov vortex. When the vortex is close to the junction, the polar angle changes stepwise, as shown in Fig.~\ref{Fig1}b, and the junction switches into the $0$~--~$\pi$ state. Yet, note that the remarkable success of the London gauge description of the phase shift around the vortex is surprising because phase shifts between any two points (such as the left and right edges of the junction) are not gauge invariant and, therefore, should not be measurable. We assume that the presence of the junction as such plays a crucial role on the way from the unmeasurable phase shift of the superconducting condensate to the measurable Josephson phase difference. Although the seeming rigidity of the gauge field around the Abrikosov vortex remains to be clarified, we demonstrate that it can be employed as a tunable and reversible phase battery for Josephson electronics. Depending on the geometrical factor $z_v/L$, such a battery can provide either a quantized step-like $\pi$-shift, or an arbitrary phase shift in the range $0<\Delta\varphi<\pi$.

\section{methods}
Planar Nb/CuNi/Nb junctions of the ``variable thickness'' type were made by cutting CuNi/Nb double layers by a Focused Ion Beam (FIB). The thicknesses of CuNi and Nb layers were $50$ and $70$~nm respectively. The junction length and width were $\sim 4$~$\mu$m and $20$~nm respectively. The details of junction fabrication and characterization can be found in Refs.~\cite{SFS,Golod}. Mesoscopic Nb/Pt$_{1-x}$Ni$_{x}$/Nb junctions of the ``overlap" type were made by 3D FIB-sculpturing. The thicknesses of bottom and top Nb layers were $225$ and $350$~nm respectively. The thickness of the Pt$_{1-x}$Ni$_{x}$ layer was varied from $20$ to $24$~nm and the junction length was varied from $110$ to $1200$~nm. To control $I_{c}$, the Ni concentration was varied from $0$ to $67$~at.\%. In Fig.~\ref{Fig4}c we show data for junctions with pure Pt barrier (circles and down triangle) and with $54$~at.\% (squares), $60$~at.\% (up triangle) and $67$~at.\% (diamond and star) of Ni. Details of junction fabrication can be found in Ref.~\cite{Golod}.
Nb, CuNi and PtNi layers were deposited by magnetron sputtering on oxidized Si wafers. The films were first patterned into $5$~$\mu$m wide bridges by photolithography and ion etching (CF$_4$ reactive ion etching for Nb and Ar-milling for CuNi and PtNi).

Measurements were done in a four probe configuration in liquid $^{4}$He or flowing gas cryostats at $T \simeq 1.8$~K. $I_{c}$ was measured by ramping the bias current and detecting the largest current corresponding to a voltage less than a certain threshold level.

\section{Acknowledgements}
We are grateful to H. Fredriksen, V.A. Oboznov and V.V. Ryazanov for assistance with sample fabrication and to T.H. Hansson and A. Karlhede for useful discussions. Financial support from the Knut and Alice Wallenberg foundation, the Swedish Research Council and the SU-core facility in nanotechnology is gratefully acknowledged.
\\
Correspondence and request for materials should be addressed to V. M. K.

\section{SUPPLEMENTARY INFORMATION}
\subsection{Controllable manipulation of the vortex by transport current}
\begin{figure}
\begin{center}
\includegraphics[width=0.90\linewidth]{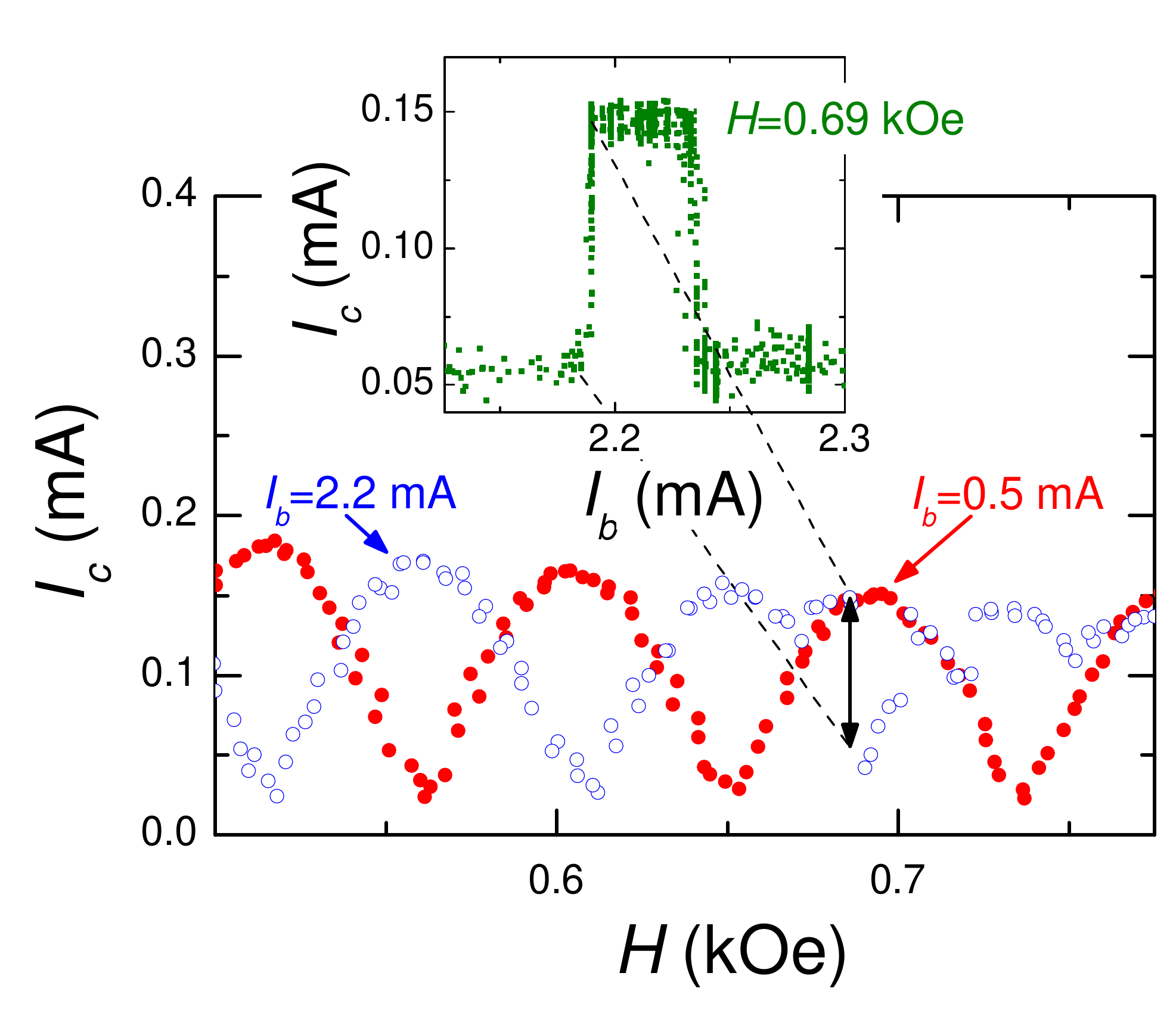}
\end{center}
\caption{{\sfbold $\bm{I_{c}(H)}$ modulations of a mesoscopic junction of size $\bm{230 \times 1140}$~nm$\bm{^{2}}$ at $\bm{T=1.8}$~K for bias currents $\bm{I_{b}=0.5}$~mA and $\bm{2.2}$~mA.} The magnetic field was applied across the $1140$~nm junction facet and swept up. {\sfbold Inset.} $I_{c}(I_{b})$ dependence at $H=0.69$~kOe. The bias current shakes the system to change the vortex state. By suddenly decreasing the current, the vortex state freezes and $I_c$ can be measured.
\label{FigS1}}
\end{figure}

Transport current exerts a Lorentz force on a trapped vortex. This allows controllable manipulation of the vortex state\cite{Finnemore}. We can remove a vortex by applying a large enough current at zero magnetic field, or introduce a vortex by applying current at finite magnetic field. The main panel in Fig.~\ref{FigS1} shows $I_{c}(H)$ for the same Nb/PtNi/Nb junction as in Fig.~\ref{Fig4}b, measured at two maximum bias currents $I_b$ (the current was swept in a saw-tooth like manner from $0$ to $I_b > I_c$). The current $I_{b}=0.5$~mA is too small to cause vortex entrance and the system stays vortex-free in the shown field range (solid symbols). However $I_b =2.2$~mA is sufficiently large to manipulate the vortex state and at $H\sim 0.7$~kOe we can switch between the vortex-free and the single Abrikosov vortex state solely by the bias current, as indicated by the arrow in Fig.~\ref{FigS1}.

Furthermore, we can controllably introduce or remove the vortex by changing the maximum bias current. The inset in Fig.~\ref{FigS1} shows $I_{c}$ as a function of $I_{b}$ at constant in-plane magnetic field $H=0.69$~kOe. It is seen that by changing $I_{b}$ from $\sim 2.1$ to $\sim 2.3$~mA the system can be switched back and forth between the two vortex states, characterized by different $I_{c}$. Our mesoscopic junctions thus act as memory cells (which can be considered as non-volatile due to the quantized nature of the vortex). The possibility to control the induced flux offset $\Delta \Phi$ in the detector junction solely by the transport current clearly demonstrates that the observed phenomenon is caused by the Abrikosov vortex.

\subsection{Analysis of $\bm{I_c(H)}$ modulation}
Assume that a single Abrikosov vortex is placed in the electrode 1 parallel to the junction plane, as shown in Fig.~1a. Supercurrent densities in electrodes 1 and 2 are determined by the second London equation:
\begin{equation}
\overrightarrow{J}_{1,2}=\frac{c}{4\pi\lambda_{1,2}^2} \left[\frac{\Phi_0}{2\pi}\overrightarrow{\nabla}\Theta_{1,2}-\overrightarrow{A}\right].
\label{London}
\end{equation}
Here $\Theta$ and $A$ are the corresponding scalar (phase) and vector potentials, and $\lambda_{1,2}$ are London penetration depths of the electrodes. The gauge-invariant Josephson phase difference $\varphi$ is obtained by integrating equation~(\ref{London}) over an infinitesimal contour of length $dx$, covering the barrier:
\begin{equation}
\frac{\partial\varphi}{\partial x}=\frac{2\pi}{\Phi_0} \left[\frac{4\pi\lambda_{1,2}^2}{c}(J_{1}^{(x)}-J_{2}^{(x)})+Bt\right].
\label{Adphi}
\end{equation}
Here $J_{1,2}^{(x)}$ are the $x$-components of the supercurrent density in the vicinity of the barrier in electrodes 1 and 2, $B$ is the in-plane ($y$-axis) magnetic induction in the barrier and $t$ is the barrier thickness.

In the Meissner state, $J_{1,2}^{(x)}$ are obtained by solving equation~(\ref{London}) in the electrodes, with boundary conditions $B = H$ outside the electrodes, $z=\pm d_{1,2}$, where $d_{1,2}$ are thicknesses of electrodes. Straightforward calculations yield\cite{Modes}:
\begin{equation}
\frac{\partial\varphi_M}{\partial x}=\frac{2\pi}{\Phi_0} \left[B \Lambda - H S \right].
\label{Adphi0}
\end{equation}
Here $\Lambda=t+\sum_{i=1,2}\lambda_i \coth\frac{d_i}{\lambda_i}$ and $S=\sum_{i=1,2} \lambda_i \textrm{cosech} \frac{d_i}{\lambda_i}$. From comparison of equation~(\ref{Adphi}) and (\ref{Adphi0}) it follows that current terms in the right-hand side of equation~(\ref{Adphi}) represent the effective magnetic flux in the electrodes, which depends on the electrode geometry, and, therefore, is not quantized. For junctions with bulk electrodes $d_{1,2} \gg \lambda$ (realized in our planar junctions) the $H$ term in equation~(\ref{Adphi0}) vanishes and one obtains a familiar expression $\partial\varphi_M / \partial x = (2\pi/\Phi_0) B (t+\lambda_1 + \lambda_2)$. For $d_{1,2} \ll \lambda$, screening by electrodes is weak and $\partial\varphi_M / \partial x \simeq (2\pi/\Phi_0) B (t+d_1/2 + d_2/2)$. Integrating $\partial\varphi/\partial x$ over the junction length $x\in [0,L]$, we obtain a simple relation between the effective flux and the total phase difference shift $\Delta \varphi =\varphi(L)-\varphi(0)$ in the junction:
\begin{equation}
\frac{\Phi}{\Phi_0}=\frac{\Delta \varphi}{2\pi},
\label{AdphiF}
\end{equation}
where $\Phi=d^*\int^L_0{B dx}$ is the effective magnetic flux in the junction and $d^*$ is the effective magnetic thickness of the junction, $d^*=t+\lambda_1 + \lambda_2$ for bulk and $d^*=t+d_1/2 + d_2/2$ for thin electrodes, respectively\cite{Modes}.

\begin{figure}
\begin{center}
\includegraphics[width=0.95\linewidth]{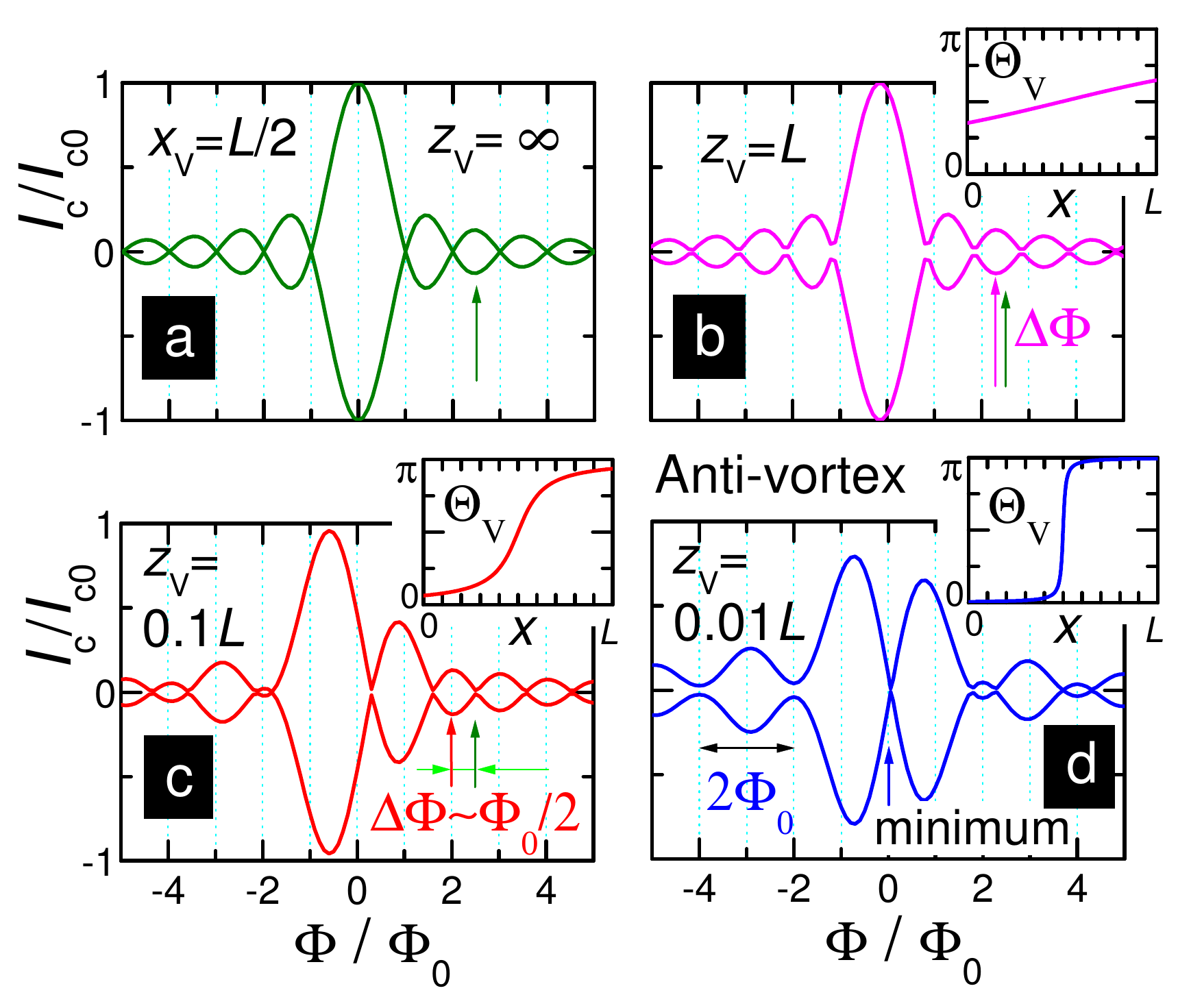}
\end{center}
\caption{{\sfbold Simulated $\bm{I_c(\Phi)}$ patterns with a single anti-vortex at different distances from the junction $\bm{z_v}$ and $\bm{x_v=L/2}$, assuming that the vortex induces the Josephson phase shift equal to the polar angle of the vortex.} Clear signatures of the $0$~--~$\pi$ junction are seen for $z_v \ll L$: The $I_c(\Phi)$ modulation becomes out-of-phase with the Fraunhofer pattern without vortex $\Delta \Phi \simeq \Phi_0/2$, as indicated by vertical arrows in panel~c. Central minimum appears at $\Phi=0$ in panel~d. Doubling of the periodicity occurs at one side of the pattern $\Phi <0$. Note that the phase gradient (induced field) has the opposite sign with respect to the vortex. {\sfbold Insets}, Vortex contribution to the Josephson phase shift corresponding to the polar angle of the vortex. For $z_v \ll L$ it turns into a sharp $\pi$-step, which causes switching of the junction into the $0$~--~$\pi$ state.
\label{FigS2}}
\end{figure}

Since the London equation~(\ref{London}) is linear, the Josephson phase difference $\varphi$ in the presence of an Abrikosov vortex in the electrode is simply given by a superposition of contributions from the Meissner state, $\varphi_M (H)$, equation~(\ref{Adphi0}), and from the vortex at zero external magnetic field, $\varphi_v(0)$:
\begin{equation}
\varphi=\varphi_M (H)+\varphi_v (0)=\varphi_M (H)+\Theta_{v1}-\Theta_{v2},
\label{varphiV}
\end{equation}
where $\Theta_{v1,2}$ are phases of the superconducting condensate in each of the electrodes, induced solely by the vortex. Within the London gauge, $\Theta_{v1}$ is simply given by the polar angle of the vortex, see Fig.~1a:
\begin{equation}
\Theta_{v1}(x)=\arctan \left(\frac{x-x_v}{z_v}\right) + \mathrm{const.}
\label{AQ1}
\end{equation}
The phase shift in the second electrode $\Theta_{v2}$ is due to magnetic flux from the vortex in the electrode-1, reaching the electrode-2. Assuming that $\Theta_{v2}$ is negligible, the Josephson phase difference caused by the vortex, $\varphi_v$, is given the polar angle, equation~(\ref{AQ1}), shown in Fig 1b and in insets of Fig.~\ref{FigS2} for different distances $z_v$ from the vortex to the junction.

The Josephson critical current in a magnetic field $H$ is given by:
\begin{equation}
I_c(H)=\frac{I_{c0}}{L}\int_0^L{\sin[\varphi(x)]dx},
\label{dcJosephson}
\end{equation}
where $I_{c0}$ is the maximum critical current at zero field. In the Meissner state, $\varphi_M(x)$ is varying linearly, with the slope proportional to the applied field, see equation~(\ref{Adphi0}): $\varphi_M(x) \simeq 2 \pi d^*B/\Phi_0 + C$, where $C$ is an arbitrary constant. The $I_c(H)$ pattern is obtain by maximization of equation~(\ref{dcJosephson}) with respect to $C$. This leads to the Fraunhofer modulation $I_{c,M}(\Phi)=I_{c0}|\sin(f)|/f$, where $f=\pi \Phi/\Phi_0$, shown in Fig.~\ref{FigS2}a.

Figures~\ref{FigS2}b-d show $I_c(\Phi)$ patterns in the presence of a single Abrikosov anti-vortex in the electrode-1, obtained by the similar maximization of equation~(\ref{dcJosephson}) with respect to the arbitrary constant. The additional contribution from the anti-vortex to the Josephson phase difference was taken to be equal to the polar angle of the anti-vortex, equation~(\ref{AQ1}), which are shown in the insets. When the anti-vortex is far away from the junction, $z_v \gg L$, the $I_c(\Phi)$ follows the conventional Fraunhofer pattern with maxima at half-integer $\Phi/\Phi_0$, as indicated by the arrow in Fig.~\ref{FigS2}a. At intermediate distances $z_v\sim L$, the anti-vortex introduces a flux offset to $I_c(\Phi)$, as indicated by the arrows in Figs.~\ref{FigS2}b,c. Note that the sign of $\Delta \Phi$ is negative, i.e., it is the same as for the introduced anti-vortex, consistent with experimental data in Figs.~2~--~4 in the main manuscript. At $z_v \ll L$, $\varphi_v$ turns into a $\pi$-step and clear fingerprints of the $0$~--~$\pi$ junction appear in the $I_c(\Phi)$, which were discussed in connection with Fig.~3b in the main manuscript.

\subsection{Discussion of other unsustainable explanations}
Phase shifts in Josephson junctions may be induced by magnetic fields and currents in the junction area. As we discussed in the manuscript, we can rule out a possibility of direct penetration of the full vortex flux $\Phi_0$ into the junction in a form of a Josephson fluxon\cite{Finnemore}. Below alternative plausible contributions are discussed in some more detail to show why they cannot explain the experimental observations either.

First we emphasize that the measured flux-offset of $I_c(\Phi)$ patterns is opposite to the trapped flux in the junction. Indeed, if a positive flux $\Phi_v$ is induced by the vortex, then it will be necessary to apply the corresponding negative field to compensate this flux. Therefore the central maximum in $I_c(\Phi)$ pattern, which correspond to zero total flux through the junction, would occur at a negative field, i.e, the whole $I_c(\Phi)$ would be shifted by
\begin{equation}
\Delta \Phi = -\Phi_v.
\label{eqnDeltaPhi}
\end{equation}

\vspace{2mm}
{\noindent \emph{Direct field of a vortex}}: Distribution of the magnetic induction in a bulk superconductor at distance $r\gg \lambda$ from the Abrikosov vortex is given by:
\begin{equation}
B(r) \simeq \frac{\Phi_0}{2\pi\lambda^2}\frac{\exp{\left(-r/\lambda \right)}}{\sqrt{r/\lambda}},
\label{eqnBr}
\end{equation}
If this field would stretch into the junction area, it would generate some Josephson phase shift, according to equation~(\ref{Adphi0}). However, this field must have the same sign as the applied field, while in experiment we always observe that the effective trapped flux is \emph{opposite} to the applied field. Thus, the phenomenon can not be due to direct field from the vortex.

\vspace{2mm}
{\noindent \emph{Effect of magnetic barrier}}: Barrier layers in our junctions contain Ni and are magnetic at low $T$ in some cases. In the Nb/Pt$_{1-x}$Ni$_x$/Nb junctions, the Ni concentration was varied from about $67 \%$ (magnetic), to pure Pt (nonmagnetic). The magnetic barrier was used to reduce the critical current to a comfortable range below about $1$~mA. The observed offset of the Fraunhofer pattern is certainly not related to the magnetism of the barrier because a ferromagnet would be magnetized in the direction of the applied field, in contrast with the experimental observations. There is no reason for the magnetic moment of a ferromagnet to induce a $\Phi_0/2$ flux contribution either. Furthermore, no differences in behavior were observed for non-magnetic Nb/Pt/Nb junctions, also displaying flux shifts. The possibility to manipulate the state by transport current clearly demonstrates that vortices in electrodes are responsible for the observed flux offsets.

\vspace{2mm}
{\noindent \emph{Vortex currents}}: Distribution of the circulating current density around the Abrikosov vortex in a bulk superconductor at distance $r\gg \lambda$ from the vortex is given by:
\begin{equation}
J_{\Theta} \simeq -\frac{c\Phi_0}{8\pi^2\lambda^3}\frac{\exp{\left(-r/\lambda \right)}}{\sqrt{r/\lambda}}\left[1+\lambda/2r \right].
\label{eqnJtheta}
\end{equation}
The influence of the vortex current on the Josephson phase shift in this case was considered theoretically in Ref.~\cite{Aslamazov}. Here we provide similar estimations for the simplified case of a junction with a cylindrical electrode of a radius $R$ with the Abrikosov vortex along the axis of the cylinder, $z_v=R$, and the detector junction covering half of the surface of the cylinder. The Josephson phase shift is obtained by integration of equation~(\ref{Adphi}) and the induced flux in the junction is obtained from equation~(\ref{AdphiF}):
%
\begin{eqnarray}
\Phi_v= \Phi_0 \frac{\Delta \varphi_v}{2\pi} \simeq \Phi_0\frac{t z_v}{2\lambda^2}\frac{\exp{\left(-z_v/\lambda \right)}}{\sqrt{z_v/\lambda}}-
\label{phivBulk} \nonumber\\
- \frac{\Phi_0}{2}\exp{\left(-z_v/\lambda \right)}\left[\sqrt{\frac{z_v}{\lambda}}+\frac{\lambda}{2z_v}\right].
\end{eqnarray}
The first positive term is due to the direct field of the vortex within the junction barrier, equation~(\ref{eqnBr}), and the second negative term is due to the vortex current at the junction interface, equation~(\ref{eqnJtheta}). If we substitute parameters for our mesoscopic Nb/PtNi/Nb junctions $\lambda \simeq 150$ nm for Nb, $t=20$ nm and $z_v\simeq 150$ nm in equation~(\ref{phivBulk}), we obtain $\Phi_v/\Phi_0 \simeq 0.25$ for the case of an anti-vortex. For the planar Nb/CuNi/Nb junctions the electrode thickness $d=70$~nm is about two times smaller than $\lambda$ and the vortex is oriented perpendicular to the thin superconducting film. In this case vortex currents decay slower (not exponentially, but quadratically) with the distance from the vortex\cite{Pearl}. On the other hand the vortex is placed substantially further away from the junction $z_v>300$~nm. In any case we see that the contribution from the vortex current may be significant. However, it depends crucially on a number of parameters: $z_v$, $\lambda$, $d$, $L$. Since we observe the same behavior for two very different types of junctions with substantially different parameters, we conclude that the observed direct correspondence of the induced Josephson phase shift to the polar angle of the vortex can not be explained by the vortex current \emph{alone}.

\vspace{2mm}
{\noindent \emph{Effect of stray fields}}: Stray fields from a vortex returning back through the junction area generate field in the direction opposite to the vortex field and would give rise to a flux shift in $I_c(\Phi)$ in the observed direction, similar to the vortex current contribution. In fact, there is a well known fundamental connection between the magnetic field outside the superconductor and the surface current in the superconductor. Therefore effects of the vortex stray field and the surface current at the junction interface are identical [follow for example the transition from equation~(\ref{Adphi}) to equation~(\ref{Adphi0})]. Discussion in terms of stray field lines from the vortex closing through the junction area is perhaps more illustrative because it shows more clearly the nontrivial behavior of the trapped flux in the junction. Indeed, in the classical case, i.e., in the absence of Josephson effect in the junction $I_c=0$, the stray field distribution should be governed by a simple magnetostatics, which minimizes magnetic field energy outside the superconductor. In this case only a small portion of the total flux should go through the junction because field focusing in the narrow slit costs extra energy. To the contrary, we observe that a half of the flux $\Delta \Phi \simeq \Phi_0/2$ is going through the junction for long enough junctions. Such quantization is not expected for the magnetostatic stray field. Most clearly the non-magnetostatic behavior of the trapped flux is demonstrated by the experiment with the stray field drain. As seen from Fig.~1c, the drain opens an additional area for returning the stray field, which is more than twenty times larger than the area of the junction slit. It is clear that this should dramatically reduce the magnetostatic flux in the junction slit. However, experimental data in Fig.~3 show that this is not the case. Apparently, magnetostatic stray field \emph{alone} can not explain the observed phenomenon.

Although the mechanism is still unclear, we believe that the presence of the Josephson junction itself plays a crucial role in the observed remarkable coincidence of the induced Josephson phase shift in the junction and the London gauge field of the vortex. Certainly, vortex stray fields and surface currents are involved. However, they are modified by the system to assume certain values through the junction. In effect, the junction creates its own magnetic flux to complement any external stray fields to give a resulting phase shift of the observed magnitude.

\end{document}